\documentstyle[version2,aps,eqsecnum,preprint]{revtex}
\begin{document}
\draft
\begin{title}

Quantum Dots in Strong Magnetic Fields:\\
Stability Criteria for the Maximum Density Droplet

\end{title}

\author{A.H. MacDonald and S.-R.Eric Yang\cite{byline}}

\begin{instit}
  Department of Physics, Indiana University, Bloomington, IN 47405
\end{instit}

\author{M.D. Johnson}
\begin{instit}
  Department of Physics, University of Central Florida, Orlando, FL
  32816-2385
\end{instit}

\begin{abstract}

In this article we discuss the ground state of a
parabolically confined quantum dots in
the limit of very strong magnetic fields where the
electron system is completely spin-polarized and all electrons
are in the lowest Landau level.  Without electron-electron
interactions the ground state is a single Slater determinant
corresponding to a droplet centered on the minimum of the
confinement potential and occupying the minimum area allowed
by the Pauli exclusion principle.
Electron-electron interactions favor droplets of larger
area.  We derive
{\it exact} criteria for the stability of the
maximum density droplet against edge excitations and
against the introduction of holes in the interior of the droplet.
The possibility of obtaining exact results in the strong
magnetic field is related to important simplifications
associated with broken time-reversal symmetry in a strong
magnetic field.

\end{abstract}


\narrowtext

\section{Introduction}

Advances in nanofabrication technology have made it
possible to realize artificial systems in which
electrons are confined to a small area within a
two-dimensional electron gas.  There
has been considerable interest in the physics of
electron-electron and electron-hole
interactions in these `quantum dot' systems\cite{qdot}.
Recent experiments have demonstrated the
possibility of probing their properties in
the regimes of the integer\cite{kastner} and
fractional\cite{hansen} quantum Hall effects\cite{qhe}.
Excited states and low-temperature
thermodynamic properties of quantum dots coupled to
particle reservoirs are discussed elsewhere\cite{ourprl}.
We focus here on the stability of the maximum-density-droplet
(MDD) state which is the ground state in the absence of
electron-electron interactions.  One interesting consequence of
the strong magnetic fields is that this state remains an
{\it exact} eigenstate of the many-particle Hamiltonian
even in the presence of electron-electron interactions.
In Section II of this paper we discuss the MDD state.
We point out that finite-size effects
in the dependence of the MDD state energy
on particle number are dominated by one-body terms from the
confinement potential rather
than by the Coulomb interactions, as is usually assumed.
In Section III we discuss the low-lying edge excitations
of the MDD state and derive a criterion for the stability
of the dot against edge excitations.  In Section IV
we consider the introduction of holes near the center of
the MDD.  We find that because of the qualitative differences
which exist between two- and three-dimensional
electrostatics the MDD becomes unstable against these
excitations before the edge becomes unstable.  Some concluding
remarks are contained in Section V.

\section{The Maximum Density Droplet }

We consider a system of electrons confined to a finite area of a two
dimensional electron gas by a parabolic potential,
$V(r) = {1 \over 2} m \Omega^2 r^2 $.
In the strong magnetic field limit where $ \Omega/\omega_c \ll 1$
only the states in the lowest Landau level are relevant.
(Here $\omega_c = e B / m c$ is the cyclotron frequency.)
In the symmetric gauge the single particle states ($\phi_l(z) \sim z^l \exp( -z
\bar z / 4 \ell^2)$)
in this level may be labeled by angular momentum
and have energy\cite{qdot}
$\epsilon_l = \hbar \omega_c/2 + \gamma (l+1)$ where
$\gamma=m\Omega^2\ell^2 = \hbar \omega_c (\Omega/\omega_c)^2 $.
(Here $z = x + i y$ is the 2D electron
coordinate expressed as a complex number,
$\ell \equiv (\hbar c / e B)^{1/2}$ is the magnetic length,
and the allowed
values of single particle angular momentum within
the lowest Landau level are $l=0,1,2,\dots$.)
For typical systems the regime where $\Omega$ is small
compared to to $\omega_c$ occurs at experimentally
available magnetic fields.
The wavefunction for the orbital with angular momentum
$l$ is localized within $\sim \ell$ of a circle of radius
$R_l$ where
$ R_l^2 \equiv  \langle l | r^2 |l  \rangle = 2 \ell^2( l+1) $.
(See Fig.~(\ref{fig:1}).)
The circle of radius $R_l$ encloses magnetic flux $(l+1) \Phi_0$,
where $\Phi_0 = hc/e$ is the electron flux quantum.
Orbitals at larger angular momentum are localized further
from the minimum of the confinement potential and
experience a stronger confinement potential.
Note that in the lowest Landau level the single-particle
orbitals all have the same sign of angular momentum.  This
consequence of broken time-reversal symmetry in the
strong magnetic field limit leads to important simplifications.
We will assume throughout this article that the magnetic
field is strong enough that mixing of states in
higher Landau levels by the electron-electron interaction
can be neglected.  We also assume that the electron system
is completely spin-polarized by the magnetic field\cite{yangspin}.

For non-interacting electrons the many-body ground state
, $|\Psi_0\rangle$,
is a single Slater determinant in which the confinement energy
is minimized by occupying orbitals
from $l=0$ to $l=N-1$.  This state is an
exact many-body eigenstate of the Hamiltonian {\it even
when electron-electron interactions are included}.
The preceding claim follows after noting that
the total angular momentum operator
\begin{equation}
\hat L_{TOT} = \sum_{l=0}^{\infty} l \hat n_l
\label{equ:1}
\end{equation}
commutes with the Hamiltonian so that $L_{TOT}$ is a
good quantum number, and that $| \Psi_0\rangle$ is the only
state in the Hilbert space with $L_{TOT}= N(N-1)/2$.
All other states have larger values of $L_{TOT}$.
However, once electron-electron interactions become important
$| \Psi_0 \rangle$ need not be the ground state.
In the lowest Landau level the total angular momentum operator
can be written
in the first quantized form
\begin{equation}
\hat L_{TOT} = \sum_i (r_i^2/2 \ell^2 -1) .
\label{equ:2}
\end{equation}
If we assume that the electrons are confined to a droplet
of roughly constant density, Eq.~(\ref{equ:2}) may be used to
relate $L_{TOT}$ to the average area of the droplet:
\begin{equation}
L_{TOT} \sim N A / (4 \pi \ell^2).
\label{equ:3}
\end{equation}
Many-body states with smaller area have smaller confinement energy
since the electrons are closer to the minimum of the
confinement potential but larger interaction energies since
the electrons are closer to each other.  For sufficiently weak
confinement the area of the ground state
of an interacting-electron droplet will increase
and $|\Psi_0 \rangle$ will no longer be the ground state.

The electron density in state $|\Psi_0 \rangle$ is:
\begin{equation}
n(r) = \sum_{l=0}^{N-1} |\phi_l(z)|^2
= \frac{1}{2 \pi \ell^2} \exp( -r^2/2 \ell^2)
\sum_{l=0}^{N-1} \frac{1}{l!} \big(\frac{r^2}{2 \ell^2}\big)^l .
\label{equ:4}
\end{equation}
Except near the edges of the droplet, $n(r) = (2 \pi \ell^2)^{-1}$.
This is the maximum electron density which can be reached at
any point without mixing states from higher Landau levels
and we therefore refer to $|\Psi_0\rangle$ as the maximum density
droplet (MDD) state.  The energy of the MDD state is
\begin{equation}
E_{\rm MDD} = \left({1\over2}\hbar\omega_c+\gamma\right)N +
\gamma N (N+1) /2 +E^H_{\rm MDD} + E^{\rm XC}_{\rm MDD}.
\label{equ:5}
\end{equation}
The first two terms on the right-hand-side of Eq.~(\ref{equ:4}) are the
kinetic energy and confinement energies.  The third term is the Hartree
(electrostatic)
energy, and the fourth term, the exchange-correlation energy, is
defined by this equation.   The Hartree energy of the MDD state
is approximately equal to that of a disk with uniform areal
number density $\bar n  = (2 \pi \ell^2 )^{-1}$:
\begin{equation}
E^H_{\rm MDD} \approx {8 e^2 N^2 \over 3 \pi \epsilon R_N} =
 {e^2 \over \epsilon \ell} { 4 \sqrt{2} \over 3 \pi} N^{3/2}.
\label{equ:6}
\end{equation}
Here $R_N = \sqrt{2N} \ell$ is the approximate radius of the
$N$-electron MDD state.
Corrections to this approximate expression for the Hartree energy
and the exchange-correlation energy will both contribute terms
$\sim N$ to the MDD state energy.

One important property of
dots which can be measured\cite{kastner,meir,been}
is the chemical potential change when a single electron is
added to the system.
If we define $\mu(N)$ as the difference
in energy between the $(N+1)$-electron ground state and the
$N$-electron ground state, then
for $ N \gg 1$ it follows from
Eq.~(\ref{equ:5}) and Eq.~(\ref{equ:6}) that
\begin{equation}
\mu(N+1)-\mu(N) =  \gamma +
{e^2 \over \epsilon \ell} { \sqrt{2} \over \pi} N^{-1/2}
\label{equ:7}
\end{equation}
up to terms vanishing as $N^{-1}$.  For a system of particles
with short-range interaction and confined to a fixed `volume'
$\Omega \sim L^d$ in d-dimensions,
$\mu(N+1) - \mu (N)$ vanishes as $N^{-1}$, leading to a
chemical potential which depends only on particle density in the
thermodynamic limit.  For ordinary small metallic grains with
$e^2 / \epsilon r$
interactions between the electrons
this quantity vanishes, at fixed density,
as $N^{-1/d}$; the anomalously slow decrease of finite-size
effects leads to the
{\it Coulomb blockade} phenomena\cite{cblock}.
We see from Eq.~(\ref{equ:7}) that for parabolically confined
two-dimensional systems in the strong magnetic
field limit the term {\it Coulomb blockade} is something
of a misnomer.   The Coulomb energy scales in the same way as for
metallic grains but the largest contribution to the
chemical potential change comes in this case
from the confinement energy.  It may be difficult to
separate these two contributions experimentally.

In the following sections we derive stability criteria
for the MDD state.  In Section III we consider the stability
of the collective phonon-like edge excitations of the MDD.
In the process we derive a useful exact identity relating Hartree-Fock
self-energies and vertex functions of the quantum dot at the
Fermi level.  This identity
is used to derive an exact expression for the
chemical potential change on adding a particle which reduces
to Eq.~(\ref{equ:7}) in the large $N$ limit.
In Section IV we consider the stability of the the MDD
against the formation of a hole in the middle of the droplet.
We find that, because of differences between two-dimensional and
three-dimensional electrostatics, with weakening confinement
the system becomes unstable to
the introduction of holes in the bulk before the edge becomes unstable.
\section{Stability of the MDD Edge}

The total angular momentum of the MDD state, $M_{\rm MDD} = N (N-1)/2 $,
is the smallest angular momentum in the Hilbert space.  The
low-energy excited states with total angular momentum
$M = M_{\rm MDD} + \delta M$ where $\delta M \ll N$
are states in which phonon-like\cite{wen,stone} collective modes
have been excited at the edge of the MDD.  In this section we
discuss the conditions required for these edge excitation
energies to be positive.  If the edge excitation energies were
not positive the MDD would not be the ground state.  This is
somewhat analogous to the soft phonon modes at wave vectors $k_{\rm F}$
and $-k_{\rm F}$ which combine to give
charge-density-wave states. Here, however, edges are chiral ($\delta M>0$)
so that if a state of nonzero $\delta M$ became the ground state the system
would retain its circular symmetry.  In fact, we present evidence that
even this does not occur, that the instability of the ground state does
not occur at the edge for parabolic quantum dots in a strong magnetic field.

The expectation value of the Hamiltonian in a single Slater determinant
state ({\it i.e.\/}, a state with definite occupation numbers
$n_m$ equal to 0 or 1) is
\begin{equation}
E[n_m] = \sum_m n_m \epsilon_m  +  {1 \over 2}
\sum_{m,m'} n_m n_m' U_{m,m'}
\label{equ:8}
\end{equation}
where
\begin{equation}
U_{m,m'} \equiv  \langle m,m' | V | m, m' \rangle -
\langle m',m | V | m, m' \rangle
\label{equ:9}
\end{equation}
is the difference of direct and exchange
two-body matrix elements.  Note that $U_{m,m}=0$.
We will show below that it is possible to express the excitation energies
for $\delta M=1$ and $\delta M=2$ in terms of such expectation values,
even though for $\delta M=2$ the eigenstates are not single Slater
determinants.
In the MDD state $n_m=1$ for $0 \le m \le N-1$ and
is zero otherwise.  Expanding the occupation numbers around
the MDD state values gives
\begin{equation}
E[n_m] = E_{\rm MDD} +
\sum_m \delta n_m (\epsilon_m + \Sigma_m^{(N)}) + {1 \over 2}
\sum_{m,m'} \delta n_m \delta n_{m'} U_{m,m'}.
\label{equ:10}
\end{equation}
Here
\begin{equation}
\Sigma_m^{(N)} = \sum_{m'=0}^{N-1} U_{m,m'}
\label{equ:11}
\end{equation}
is the Hartree-Fock self-energy for the $N$-electron
MDD state and $\epsilon_m + \Sigma_m^{(N)}$ is the Hartree-Fock
quasiparticle energy.
The Hartree-Fock self-energy is shown in Fig.~(\ref{fig:2})
for an $N=40$ MDD state.  We will see below that because of the broken time
reversal symmetry some properties of the system's excitations
are given exactly in terms of the Hartree-Fock self energy.

We first consider the state with $\delta M=1$.
There is only one state in the Hilbert space at this angular
momentum and it is therefore an exact eigenstate of the
Hamiltonian.  This state, which we label $|1\rangle$,
has $\delta n_N=1$ and $\delta n_{N-1} = -1$
as illustrated in Fig.~(\ref{fig:3}).   From Eq.~(\ref{equ:10}) it
follows that
\begin{equation}
E_1 = E_{\rm MDD} + \gamma + \Sigma_N^{(N)} - \Sigma_{N-1}^{(N)}
-U_{N,N-1}.
\label{equ:12}
\end{equation}
(In a perturbative treatment the contribution $U_{N,N-1}$ to the
excitation energy would appear through vertex corrections
to a two-particle Greens function.)
However, $|1\rangle$ differs from the MDD state only through
an excitation of the center-of-mass.
To see this it is convenient to define a first-quantized ladder operator
for center of mass states in the lowest Landau level:
\begin{equation}
B^{\dagger} = {1 \over \sqrt{N}} \sum_i b_i^{\dagger}.
\label{equ:13}
\end{equation}
Here $b_i^{\dagger}=(z_i/2\ell -2\ell \partial/\partial\bar z_i)/\sqrt{2}$
is the intra-Landau-level single-particle ladder
operator\cite{qhereview} which can be used to generate the
angular momentum eigenstates in the Landau gauge
($b^{\dagger} | m \rangle = \sqrt{m+1}  | m+1 \rangle$).  The
center-of-mass states in the lowest Landau level have the same
set of angular momenta as the single-particle states and are
generated from the zero angular momentum center-of-mass state
by $B^{\dagger}$. In second-quantized form $B^{\dagger}$ and the center-of-mass
angular momentum operator $M_{\rm COM}=B^{\dagger}B$ can be written
\begin{eqnarray}
&\displaystyle \hat{B}^{\dagger} = {1 \over \sqrt{N}} \sum_m
 \sqrt{m+1} \hat c^{\dagger}_{m+1} \hat c_m,  \label{equ:14}\\
&\displaystyle \hat{M}_{COM} = { 1 \over N} \left[
\sum_{m>0} \hat mn_m + \sum_{m,m'} \sqrt{(m+1)m'} \hat c_{m+1}^{\dagger}
\hat c_{m'-1} ^{\dagger}\hat c_{m'} \hat c_m \right] .
\label{equ:14a}
\end{eqnarray}
Here $\hat c_m^{\dagger}$ creates an electron in the single-particle state
$\phi_m$ in the lowest Landau level.
$\hat B$ and $\hat B^{\dagger}$ obey boson commutation relations:
$[\hat B,\hat B^{\dagger}] = \sum_m \hat n_m/N$, which is unity in the
$N$-electron sector.
It is easy to verify that $|MDD\rangle$ is an eigenstate of
$\hat M_{COM}$ with eigenvalue zero and that
$B^{\dagger} |MDD\rangle = |1\rangle$.
Since $B^{\dagger}$ operates only on the center-of-mass
degree of freedom it commutes with the interaction part of
the Hamiltonian\cite{trugman} and
\begin{equation}
[H,B^{\dagger}]=\gamma B^{\dagger}.
\label{equ:15}
\end{equation}
It follows that $E_1 = E_{\rm MDD} + \gamma$.  Note that
$|1\rangle$ has a higher energy than $|MDD\rangle$ no
matter how weak the confinement.
Comparing with
Eq.~(\ref{equ:12}) it follows that
\begin{equation}
\Sigma_N^{(N)} = \Sigma_{N-1}^{(N)} + U_{N,N-1}.
\label{equ:16}
\end{equation}
This exact relationship between the self-energy and the vertex
correction is a consequence of the fact that the relative motion
in $|MDD\rangle$ and $|1\rangle$ is identical.  We will use this
relationship below to calculate the energies of the $\delta M =2$
states.

This approach to generating edge excitations has been used for other
purposes.
In first quantized language, a bosonic basis for the edge excitations
can be constructed by the power sums\cite{stone} $S_k=\sum_i z_i^k$.
The $M=M_{\rm MDD}+\delta M$ subspace is spanned by the
set of products $\{S_1^{l_1}S_2^{l_2}S_3^{l_3}\dots\}|MDD\rangle$ with
$\sum_k k l_k=\delta M$.  The operator $B^{\dagger}$ given in
Eq.~(\ref{equ:13})
is, when acting on states in the lowest Landau level, equivalent to $S_1$ times
a normalization constant.  ($B^{\dagger}$ is written in terms of
$b_i^{\dagger}$
while $S_1$ is written in terms of $z_i$.  The benefit of the former is that
its adjoint is easy to determine.)  The operator $\hat B^{\dagger}$ is
thus the second-quantized form of $S_1$, and second-quantized versions of
the remaining $S_k$ have also been constructed\cite{marsili}:
within a normalization constant,
$\hat B_k^{\dagger} = \sum_m \sqrt{(m+k)!/m!} \hat c^{\dagger}_{m+k} \hat
c^{\phantom{\dagger}}_m$.
Only for $k=1$ (the case discussed above) does the operator
$\hat B^{\dagger}_k$ generate an eigenstate for finite $N$.

There are two states in the many-body Hilbert space with
$\delta M =2$; one has $\delta n_{N-2}=-1$ and $\delta n_{N}=1$
and is labeled as state $|2:A\rangle$ in Fig.~(\ref{fig:3})
while the other has $\delta n_{N-1} = -1$ and $\delta n_{N+1} = 1$
and is labeled $|2:B\rangle$ in Fig.~(\ref{fig:3}).  We can easily
generate one of the two eigenstates at $\delta M=2$ by using
the center-of-mass angular-momentum raising operator:
\begin{equation}
|2;+\rangle \equiv {1 \over \sqrt{2} } (B^{\dagger})^2 |MDD\rangle.
\label{equ:17}
\end{equation}
It follows from Eq.~(\ref{equ:15}) that $|2;+\rangle$ is an eigenstate
of the Hamiltonian with eigenvalue $E_{2+} = E_{\rm MDD} + 2 \gamma$.
Applying $\hat B^{\dagger}$ twice, we see that
\begin{equation}
|2;+\rangle = \big( {N-1 \over 2N} \big)^{1/2} |A\rangle
+ \big( {N+1 \over 2N} \big)^{1/2} |B\rangle.
\label{equ:19}
\end{equation}
The other eigenstate at $\delta M=2$ must be orthogonal to this, so
\begin{equation}
|2;-\rangle = \big( {N+1 \over 2N} \big)^{1/2} |A \rangle
- \big( {N-1 \over 2N} \big)^{1/2} | B \rangle.
\label{equ:20}
\end{equation}
(This state can be written as a linear combination
$(\alpha \hat B_2^{\dagger} + \beta (\hat B^{\dagger})^2)|MDD\rangle$.
In the limit $N\rightarrow\infty$, $\beta\rightarrow0$.)

It follows that
the eigenenergy of this state is
$E_{2-} = E_{\rm MDD} + 2 \gamma  + \delta E$ where
\begin{equation}
\delta E = 2 N [\langle A | \hat V | A \rangle - E_{\rm MDD}^{\rm int}] / (N+1)
=
2 N [ \langle B | \hat V | B \rangle - E_{\rm MDD}^{\rm int} ]  / (N-1).
\label{equ:21}
\end{equation}
(Here $E_{\rm MDD}^{\rm int}=E_{\rm MDD}^H + E_{\rm MDD}^{XC}$ is the
interaction energy in the MDD state.)
This can be shown in two steps.  First, calculate the expected energies
of $|2;+\rangle$ and $|2;-\rangle$ and use the known value of the
former to eliminate the off-diagonal matrix element $\langle A|\hat
V|B\rangle$.
Second, solve the $2\times2$ Hamiltonian with basis states $|A\rangle$ and
$|B\rangle$, and require that the resulting eigenstates be
Eq.~(\ref{equ:19}) and Eq.~(\ref{equ:20}).   This gives Eq.~(\ref{equ:21}).
Also, from Eq.~(\ref{equ:10}), Eq.~(\ref{equ:11}) and Eq.~(\ref{equ:16})
it follows that
\begin{equation}
\langle A | \hat V | A \rangle  = E_{\rm MDD}^{\rm int} + U_{N,N-1} -
U_{N,N-2}.
\label{equ:22}
\end{equation}
We have thus succeeded in expressing the eigenenergies for
$\delta M =2$ in terms of interaction matrix elements near
the edge of the dot.

For large $N$ the above results may be used to obtain a necessary
condition for the stability of the maximum density droplet.
As illustrated in
Fig.~(\ref{fig:1}) the single-particle orbital with
angular momentum $N$ is localized
within about $\ell$ of a circle of radius $R_N = \sqrt{2N} \ell$.
If we ignore the width of the resulting annulus in comparison
with its circumference, an approximation which becomes
increasingly accurate as $N$ increases, we obtain, for $M\sim N$,
\begin{equation}
U_{N,M}  \approx {e^2 \over 4\pi \epsilon R_N} \int_{0}^{2 \pi}
{ 1 - \cos[(N-M)\theta] \over \sin(\theta/2) } d\theta
= {2 e^2 \over \epsilon R_N \pi} \sum_{l=1}^{|N-M|} {1 \over 2 l-1}.
\label{equ:23}
\end{equation}
The second term in the numerator of the integrand for the integral
over $\theta$ comes from the exchange term.  If this term were
not present $U_{N,N\pm k}$ ($ k \ll N$) would be logarithmically
larger for large $N$:
$U_{N,N\pm k} \sim (e^2 \ln (R_N / \ell))/(\epsilon 2R_N)$.
Comparing Eq.~(\ref{equ:23}), Eq.~(\ref{equ:22}) and Eq.~(\ref{equ:21}),
we see that for large $N$
\begin{equation}
E_{2-} - E_{\rm MDD} =  2 \gamma - {4 e^2  \over 3 \pi \epsilon R_N}.
\label{equ:24}
\end{equation}
The interaction energy is lowered in this state because the electrons
are spread over a slightly larger area.  In $|2;+\rangle$, on the
other hand, the center of mass of the droplet is not as well
localized but the area of the droplet stays the same.
We can conclude from the above exact result that the MDD state
becomes unstable at the edge if
\begin{equation}
{\gamma \over e^2 / \epsilon \ell } < {\sqrt{2} \over 3 \pi N^{1/2}} = 0.15005
N^{-1/2}.
\label{equ:25}
\end{equation}

We should now consider the possibility that the edge instability
occurs first for larger $\delta M$.  For $\delta M = 3 $ there
are three states, two of which we can easily generate using
the center-of-mass angular momentum raising operator:
$B^{\dagger}|2,+\rangle/\sqrt{3}$ and $B^{\dagger}|2;-\rangle$.
These two states have energies larger by $\gamma$ than $|2;+\rangle$
and $2,-\rangle$, respectively, and are always more stable than
states already considered.  The third eigenstate is the one
orthogonal to these two and its energy could be evaluated using
the same approach as above.  It has all of its excess angular momentum
in the relative motion of the electrons and should be the lowest
energy $\delta M =3 $ state.  Although we have not yet completed
this calculation we expect that the third state becomes unstable
before $|2;-\rangle$.  However as
we show in the following section, it becomes energetically
favorable to introduce holes in the bulk of the
MDD well before the edge of the MDD becomes unstable.
The first instability occurs at $\delta M=\sim N$ for an
$N$-electron droplet and does not correspond to an edge
excitation.

The results in this section can be used to derive a simple
exact expression for the chemical potential change on
adding a particle.
Defining $\mu(N)$ as the difference
in energy between the $N+1$ electron ground state and the
$N$-electron ground state, as earlier,
it follows from Eq.~(\ref{equ:10})
that $\mu(N) = \gamma N + \Sigma_N^{(N)}$.   Using
Eq.~(\ref{equ:11}) and Eq.~(\ref{equ:16}) it then
follows that
\begin{equation}
\mu(N+1)-\mu(N) = \gamma + U_{N+1,N}.
\label{equ:25a}
\end{equation}
Using Eq.~(\ref{equ:23}) for $U_{N+1,N}$ at large $N$
we recover the results of Eq.~(\ref{equ:7}).
We emphasize that even though these results are expressed
in terms of Hartree-Fock approximations they are in fact
exact as long as the MDD state remains the ground state.

\section{Bulk Hole Instability}

In this section we consider the single Slater determinant
$|1H\rangle$ which differs from the maximum density droplet
by having $\delta n_0 = -1$ and $\delta n_{N}=1$.  This state
differs from the states $|1\rangle$ and $|2:A\rangle$ only
in that the orbital which is emptied is at the center of
the droplet.  The state can be considered as an $N+1$ electron
droplet with a hole at the center.  Unlike the cases discussed
above there are many $N$-particle states of the droplet
with the same total angular momentum as $|1H\rangle$.
However the coupling between $|1H\rangle$ and the other states
is weak for large droplets and we will ignore it in the
discussion below.  (The states with the same angular momentum
as $|1H\rangle$ have the hole in a state of angular momentum
$m$ and the edge of the $N+1$ electron droplet in a state
with the same excess angular momentum.  The coupling matrix
elements can be shown to scale as $e^2 \ell^m/ \epsilon R_N^{m+1}$.)
Using Eq.~(\ref{equ:10}) we see that the energy of the hole state is
\begin{equation}
E_{1H} = E_{\rm MDD} + N \gamma + \Sigma_{N}^{(N)} - \Sigma_0^{(N)}
-U_{0,N}.
\label{equ:26}
\end{equation}
The last term represents an excitonic attraction between the
hole and the extra charge at the edge, is proportional to
$N^{-1/2}$, and becomes negligible for large dots.
The Hartree contribution to the self-energies can be estimated
from the Hartree potential of a uniformly charged droplet of
radius $R_N$ as discussed in Section II:
\begin{equation}
V_H^{(N)}(r) = {2 N e^2  \over \epsilon R_N} F({1 \over 2},-{1 \over 2}:1,{r^2
\over R^2}).
\label{equ:27}
\end{equation}
$V_H^{(N)}(r)$ is plotted in Fig.~(\ref{fig:4}).  Note that
it is a monotonically decreasing potential, unlike the potential
from a uniformly charged sphere in three dimensions which
increases monotonically with radius inside the sphere.
The different behavior can be related to the larger fraction
of charge inside a given radius in the two-dimensional case.
This difference between two-dimensional and three-dimensional
electrostatics plays a very important role in determining
the properties of quantum dots, particularly in the strong
magnetic field limit as we see below.

Note that the Hartree potential is proportional to $N^{1/2}$ so
that for large droplets the exchange contribution to the self-energy
is negligible by comparison.   Taking
$\Sigma_{N}^{(N)}- \Sigma_0^{N} \approx V_H^{(N)} (R_N) - V_H^{(N)}(0) =
{ N e^2  \over \epsilon R_N} (4/\pi - 2)$, we see that for
large dots it
becomes favorable to introduce holes at the center of the
MDD state whenever
\begin{equation}
{\gamma \over e^2/ \epsilon \ell } <
{\sqrt{2} - \sqrt{8}/\pi \over N^{1/2}} = 0.51390 N^{-1/2}.
\label{equ:28}
\end{equation}
Expanding Eq.~(\ref{equ:27}) near the center of the droplet we
see that the sum of confinement and Hartree potentials is given by
\begin{equation}
V_{CH}(r) = {e^2 \sqrt{2N} \over \epsilon \ell}
+ \left({r \over R_N}\right)^2 \left( \gamma N - {N^{1/2} e^2
\over \sqrt{8} \epsilon \ell} \right).
\label{equ:28a}
\end{equation}
At the point where it becomes favorable to introduce holes
at the center of the droplet the quasiparticle energy will
increase with angular momentum.  It follows that holes
will be first introduced in the bulk away from the center of
the droplet and at a slightly larger value of $\gamma$ than
required for the introduction of holes at the
center of the droplet.  Detailed results, given elsewhere\cite{yang93},
depend on the number of electrons in the droplet and require
numerical calculations.

The introduction of holes in the bulk of the dot preempts
the edge stability discussed in the previous section.  The
holes in the bulk of the droplet increase the strength
of the confinement field seen at the edge of the disk
and prevent the edge from becoming unstable as the density is lowered
further.  In Fig.~(\ref{fig:5}) we plot the Hartree-Fock
quasiparticle energies for an $N=40$ MDD state including a
single-particle contribution for
$\gamma = 0.07 (e^2 / \epsilon \ell)$.  This value
is just large enough to ensure stability and may be
compared with the critical gamma for holes at the center
of the droplet from the above large $N$ approximation which gives
$\gamma \sim 0.08 (e^2 / \epsilon \ell)$.
Note that for this size droplet the
instability will occur first for $m \sim 15$ corresponding
to $\delta M \sim 25$.  Also plotted in Fig.~(\ref{fig:5})
are the quasiparticle
energies obtained neglecting the exchange contribution.  We
see that the MDD state is already unstable if exchange is
neglected.  The Hartree approximation seriously underestimates
the stability of the MDD state and would lead to qualitatively
incorrect results.

\section{Summary and Concluding Remarks}

For non-interacting electrons the ground state of a parabolically
confined $N$-electron quantum dot at strong magnetic fields has the
single-particle orbitals with $m= 0, 1, \cdots , N-2,N-1$ occupied.
Because of the strongly broken time-reversal symmetry at
strong magnetic fields this maximum-density-droplet state remains
an exact eigenstate of the Hamiltonian including electron-electron
interactions.  In this paper we have examined the conditions required
for the MDD to remain the ground state.  We have found that for
large dots the MDD is unstable toward edge excitations
for $ \gamma / (e^2 / \epsilon \ell ) <  0.15005 N^{-1/2}$ where
$\gamma$ is a parameter which measures the strength of the
confinement potential.  However for
$\gamma / (e^2/ \epsilon \ell) < 0.51390 N^{-1/2}$ we find that
the MDD is unstable toward the introduction of holes at the
center of the system.  This behavior is directly related to
differences between two-dimensional and three-dimensional
electrostatics.
The critical value of $\gamma $ at which holes are introduced
in the lowest Landau level will, at least for large droplets,
will be approximately equal to the value of $\gamma$ at which
the spins first become fully spin-polarized\cite{yangspin}.
These small values of $\gamma$ are those at which the
fractional Hall regime is first being approached in quantum
dots.  For quantum dots in GaAs, $\gamma \sim
0.58 (\hbar \Omega [{\rm meV}])^2 / B [{\rm Tesla}]).$
Thus it seems that this regime can be reached with magnetic
fields available in the laboratory.

One of us (AHM) acknowledges the hospitality and support of the
school of physics at the University of New South Wales during
the period when this article sprang to life.
This work was supported in part by the National Science Foundation
under grant DMR-9113911 and in part by the UCF Division of Sponsored
Research.

\newpage

\figure{Schematic representation of lowest Landau level
orbits for a quantum dot in a magnetic field.  With
increasing radii, each circle encloses an additional
unit of area and represents an orbital with an additional
unit of angular momentum.   Higher angular momentum
orbitals are farther from the minimum of the confinement
potential and have larger confinement energy.  In
an $N$-electron maximum-density-droplet state the
innermost $N$ orbitals are occupied and others are empty.
In this illustration the first twenty circles (solid lines)
represent orbitals occupied in a twenty-electron
maximum-density-droplet state, and the dashed circles represent
unoccupied orbitals.
\label{fig:1}}

\figure{Hartree-Fock self-energy [Eq.~\ref{equ:11}] for $m=0$ to $m=50$ for the
maximum-density-droplet with $N=40$.  The diamonds show the
Hartree self-energy, which neglects exchange, the
squares show the full self-energy, and the crosses show
the negative of the exchange energy.
\label{fig:2}}

\figure{Occupation numbers for the states considered in this
section.  Occupied states are indicated by solid circles and
unoccupied states by open circles.
\label{fig:3}}

\figure{Hartree potential from a uniformly charged disk of
radius $R$.  Unlike the three-dimensional case the potential
is larger at the center of the disk than at the edge of the
disk.  The dashed line shows the potential when the charge
is collapsed to a point at the center of the disk.
\label{fig:4}}

\figure{Hartree-Fock quasiparticle energies for $m=0$ to $m=50$
(squares) for $N=40$
and $\gamma = 0.07 e^2/ \epsilon \ell$.  Note that the
occupied orbitals all have lower energy than any
unoccupied orbital.  The proximity of the bulk hole
instability is evident.  The diamonds show the quasiparticle
energies in the Hartree approximation where exchange is neglected.
\label{fig:5}}


\begin{thebibliography}{10}

\bibitem[*]{byline} Permanent address: IMS, NRC of Canada,
Ottawa K1A 0R6, Canada.

\bibitem{cblock} H.R. Zeller and I. Giaver, Phys. Rev.
{\bf 181}, 789 (1969); D.V. Averin and K.K. Likharev in
{\it Mesoscopic Phenomena in Solids}, edited by B.L. Altschuler,
P.A. Lee, and R.A. Webb (Elsvier, Amsterdam, 1991) p. 173.

\bibitem{maksym} P. A. Maksym and Tapash Chakraborty,
Phys. Rev. Lett. {\bf 65}, 108 (1990).

\bibitem{ourprl} A.H.MacDonald and M.D. Johnson, submitted
to Phys. Rev. Lett. (1992).

\bibitem{qhe} K. von Klitzing, G. Dorda, and M. Pepper,
Phys. Rev. Lett. {\bf 45}, 494 (1980);
D.C. Tsui, H.L. Stormer, and A.C. Gossard,
Phys. Rev. Lett. {\bf 48}, 1559 (1982).

\bibitem{hansen} W. Hansen, T.P. Smith III, K.Y. Lee, J.A. Brum,
C.M. Knoedler, J.M. Hong, and D.P. Kern, Phys. Rev. Lett.
{\bf 62}, 2168 (1989).

\bibitem{qdot}  For recent reviews see
U. Merkt, Advances in Solid State Physics, {\bf
30}, 77 (1990); Tapash Chakraborty,
Comments on Condensed Matter Physics {\bf 16}, 35 (1992);
M.A. Kastner, Rev. Mod. Phys. {\bf 64}, 849 (1992).

\bibitem{yangspin}  For a discussion of the situation where
the spin degree-of-freedom becomes important see S.-R.Eric Yang,
A.H. MacDonald, and M.D. Johnson, Phys. Rev. B (submitted
for publication, 1992).

\bibitem{kastner} P.L. McEuen, E.B. Foxman, U. Meirav,
M.A. Kastner, Y. Meir, Ned S. Wingreen, and S.J.
Wind,
Phys. Rev. Lett. {\bf 66}, 1926 (1991); P.L. McEuen, E.B. Foxman, Jari Kinaret,
U. Meirav, M.A. Kastner, Ned S. Wingreen, and S.J. Wind,
Phys. Rev. B {\bf 45}, 11419 (1992).

\bibitem{meir} Y. Meir, N.S. Wingreen, and P.A. Lee, Phys.
Rev. Lett. {\bf 66}, 3048 (1991).

\bibitem{been} C.W.J. Beenakker, Phys. Rev. B {\bf 44}, 1646 (1991).

\bibitem{wen} X.G. Wen, review article
submitted to {\it Int. J. Mod. Phys. B\/}, and references therein.

\bibitem{trugman} S.A. Trugman and S. Kivelson, {\it Phys. Rev. B\/} {\bf 31},
5280 (1985).

\bibitem{stone} Michael Stone, {\it Phys. Rev. B\/} {\bf 42} 8299, (1990);
{\it Ann. Phys. (NY)\/} {\bf 207}, 38 (1991);
{\it Int. J. Mod. Phys. B\/} {\bf 5}, 509 (1991);
Michael Stone, H.W. Wyld, and R.L. Schult,
{\it Phys. Rev. B\/} {\bf 45} 14156 (1992).

\bibitem{marsili} M. Marsili and E. Tosatti, preprint.

\bibitem{qhereview} A.H. MacDonald in {\it Quantum Coherence in
Mesoscopic Systems}, edited by B. Kramer (Plenum, New York, 1991)
p. 195.

\bibitem{yang93} S.-R. Eric Yang, M.D. Johnson and A.H. MacDonald,
to be submitted for publication (1993).

\end{thebibliography}
\end{document}